\documentstyle[12pt]{article}\newcommand{\nc}{\newcommand}  \nc{\ov}{\over}
\nc{\hf}{{1\ov2}} \nc{\ra}{\rightarrow} \nc{\iy}{\infty}
\nc{\bq}{\begin{equation}} \nc{\eq}{\end{equation}} \nc{\ph}{\varphi}
\nc{\bZ}{{\bf Z}} \nc{\bC}{{\bf C}} \nc{\la}{\lambda} \nc{\Ga}{\Gamma}
\nc{\inv}{^{-1}} \nc{\om}{\omega} \nc{\al}{\alpha} \nc{\noi}{\noindent}
\renewcommand{\sp}{\vspace{1ex}}  \nc{\be}{\beta} \textwidth=6.5in
\begin{document}  \begin{center}{\large\bf Asymptotics of a Class of Fredholm
Determinants}\end{center} \sp\begin{center}{{\bf Craig A. Tracy}\\ {\it
Department of Mathematics and Institute of Theoretical Dynamics\\ University of
California, Davis, CA 95616, USA\\ e-mail address:
tracy@itd.ucdavis.edu}}\end{center} \begin{center}{{\bf Harold Widom}\\ {\it
Department of Mathematics\\ University of California, Santa Cruz, CA 95064,
USA\\ e-mail address: widom@math.ucsc.edu}}\end{center}\sp

\begin{abstract} In this expository article we describe the asymptotics
of certain Fredholm determinants which provide solutions to the
cylindrical Toda equations, and we explain how these asymptotics are
derived. The connection with Fredholm determinants  arising in the theory
of random matrices, and their asymptotics, are also discussed. 
\end{abstract}\sp

\noi{\bf 1. Introduction}\sp

In the Gaussian unitary ensemble, scaled in the bulk so that the density
of  eigenvalues becomes one, the probability $E(0;s)$ that an interval of
length $s$ contains no eigenvalues is equal to the Fredholm determinant
of the operator on $L^2(0,s)$ with kernel 
\bq{1\ov\pi}\,{\sin\pi (x-y)\ov
x-y}.\label{sinkern}\eq 
This is a result of M.~Gaudin \cite{G}. This
quantity is directly related to the spacings  between consecutive
eigenvalues. The conditional probability density that, given that there
is an eigenvalue at $x$, the next larger eigenvalue lies in an
infinitesimal neighborhood of $x+s$ is equal to $-d^2E(0;s)/ds^2$. This quantity is 
independent of $x$ since the kernel depends only on the difference of the arguments.

To see how large the gaps between successive eigenvalues might be the
asymptotics of $E(0;s)$ as $s\ra\iy$ become of interest. (The asymptotics
as $s\ra0$ are  trivial.) A complete asymptotic expansion was found by
F.~J.~Dyson \cite{D}, and reads
\[E(0;s/\pi)\sim 2^{1/3}\,e^{3\zeta'(-1)}\,s^{-1/4}\,e^{-s^2/8}\, \left(1+{a_1\ov
s}+{a_2\ov s^2}+\cdots\right),\]
with computable constants $a_1,\ a_2,\cdots$. The main part of the
asymptotics  (all but the parenthetical correction terms) were obtained
by applying a formal scaling argument to a result of the second author
\cite{W1} on the aymptotics of certain Toeplitz determinants. The
asymptotics were completed using inverse scattering techniques.

The derivation described above was not rigorous. The first rigorous
result of this type is also due to the second author \cite{W2,W3} and was
that $E(0;s)=e^{-s^2/8+o(s)}$. More exactly, \[{d\ov ds}\log
E(0;s)=-{s\ov4}+o(1).\] This was subsequently extended to a complete
asymptotic expansion by Deift, Its and Zhou \cite{DIZ}. Thus all of the
Dyson expansion has now been proved, except for the  constant factor. 

Incidentally this constant is expressible in terms of the Barnes
$G$-function \cite{B}, which will also arise below. This is an entire
function satisfying $G(1)=1$ and the functional equation
\[G(z+1)=\Ga(z)\,G(z).\] 
It turns out that 
\[G(\hf)^2=2^{1/12}\,\pi^{-1/2}\,e^{3\zeta'(-1)}.\]

The kernel (\ref{sinkern}) arising in the Gaussian unitary ensemble has
the form  
\bq{\ph(x)\,\psi(y)-\psi(x)\,\ph(y)\ov x-y},\label{kernform}\eq
with $\ph=\sin\pi x/\sqrt{\pi}$ and $\psi(x)=\cos\pi x\sqrt{\pi}$. Many
other kernels  of this same form arise in random matrix theory and for
some there are (heuristically  derived) asymptotic expansions analogous
to the Dyson expansion \cite{TW1,TW2}. One of the interesting things
about the function $E(0;s)$ is that it is expressible in terms of a
Painlev\'e transcendent, more precisely $P_V$ \cite{JMMS}. It was shown
by the authors \cite{TW1,TW2,TW3} that analogous quantities  arising in
several other matrix ensembles are also expressible in terms of
Painlev\'e functions. All the kernels had the same general form
(\ref{kernform}), and this fact was crucial for the argument. 

The first connection between Fredholm determinants and a Painlev\'e function was
established in a paper of B.~M.~McCoy, T.~T.~Wu and the first
author \cite{MTW}. The function is the solution to the
cylindrical sinh-Gordon equation  (reducible to a special case of the
$P_{III}$ equation)
\bq q''(t)+{1\ov t}\,q'(t)=8\,\sinh 2q\label{sG}\eq
which satisfies
\[q(t)\sim {\la\ov\sqrt{2\pi t}}\,e^{-4t}\ \ \  {\rm as}\ t\ra\iy.\]
In the cited work $q$ was shown to be expressible as a certain sum of multiple integrals.
This sum is in fact equal to a combination of logarithms of Fredholm determinants
of two integral operators. Soon thereafter \cite{SMJ} an integral representation was found,
in terms of this same function $q$, for $\det\,(I-\la^2\,K^2)$, where $K$ is the
integral operator on $L^2(0,\infty)$ with kernel
\bq {e^{-t\,(x+x\inv+y+y\inv)}\ov x+y}.\label{K}\eq
This is also of the form
(\ref{kernform}). (More precisely it can be brought to that form with the
change of variable $x\ra x^{1/2}$.) More recently \cite{BLC,TW4} it was discovered that 
there is a direct connection between this function and Fredholm determinants
associated with $K$. This is $q=\log\,\det(I+\la K)-\log\,\det(I-\la K)$. (With hindsight
it can be seen that this representation is equivalent to the one found in \cite{MTW}.)
It is a generalization of the kernel (\ref{K}) which we shall be considering here.

The {\it connection problem} for equations such as (\ref{sG}) is to determine
the behavior of a singular point (in this case $t=0$)  if the behavior at
another singular point (in this case $t=\iy$) is known. This connection
problem was solved in \cite{MTW}; the asymptotics will be a special case
of those we shall describe below. The Dyson expansion, of course, solves
the connection problem for another Painlev\'e function.

We shall describe here the results and methods of \cite{TW5} which give
the  asymptotics of the Fredholm determinants of a class of kernels
containing the above $K$ as a special case. These determinants provide
solutions to the cylindrical Toda equations
\bq q_k''(t)+{1\ov t}\,q_k'(t)=4\,(e^{q_k-q_{k-1}}-e^{q_{k+1}-q_{k}}),
\ \ \ \ \ k\in\bZ.
\label{T}\eq
If we have a solution $q$ of (\ref{sG}) then $q_k=(-1)^k\,q$ is a
solution to (\ref{T}) which is 2-periodic in the sense that $q_{k+2}=q_k$
for all $k$. Conversely any 2-periodic solution of (\ref{T}) gives the
solutions $q=(q_{k} -q_{k-1})/2$ of (\ref{sG}). 

We introduce a class of $n$-periodic solutions to the Toda equations
generalizing the solution stated above for the sinh-Gordon equation.
Define a family of kernels indexed by $k\in\bZ$ by the formulas
\[K_k(x,y)=
\sum_{\stackrel{\om^n=1}{\om\neq1}}c_{\om}\,\om^k\,
{e^{-t\,[(1-\om)\,x+(1-\om\inv)\,x\inv]}\ov -\om x+y}.\]
This is clearly $n$-periodic in $k$, and it was shown in \cite{W4} that
\bq q_k=\log\,\det(I-K_k)-\log\,\det(I-K_{k-1})\label{qk}\eq
provides a solution to (\ref{T}). In the special case $n=2$ there is a
single coefficient and we recover the solution to (\ref{sG}) described
above. (The exponential factors in the kernels are different, but the
determinants are easily shown to be the same.)

It is the asymptotics as $t\ra0$ of these determinants $\det(I-K_k)$
which we shall discuss,  and of course this will give the asymptotics of
the solution $q_k$. The  asymptotics take the form
\[\det(I-K_k)\sim b\,t^a,\ \ \ \ \ (t\ra0),\]
where the constants $a$ and $b$ are determined by $k$ and the
coefficients $c_{\om}$. (We give the formulas for them and state our
precise assumptions below.)

Asymptotic formulas of this type in the $n=2$ case for $\det(I-K^2)$ were obtained
earlier. The constant $a$ was determined in $\cite{J}$ and $b$ in \cite{BT}. 
In fact these
were obtained for a more general class of kernels, where $K^2$ is replaced by $KK^t$ 
and $K$ now denotes the kernel (\ref{K}) multiplied by a factor $(x/y)^{\theta}$. As 
was already mentioned, the asymptotics of $q$ for the $n=2$ case were found in 
\cite{MTW}; in \cite{K} a method was described to find connection formulas for 
solutions of a class of equations including some of the $n=3$ cases of (\ref{T}).\newpage

\noi{\bf 2. Statement of the results}\sp

Since $K_k$ is obtained from $K_0$ by multiplying the coefficients $c_{\om}$ by 
$\om^k$ we may as well assume that $k=0$. We do this and write $K$ instead of $K_0$. It 
is also convenient to change the constant $b$ so that the asymptotic formula reads
\[\det(I-K)\sim b\,\Big({t\ov n}\Big)^a,\ \ \ \ \ (t\ra0).\]
The constants $a$ and $b$ are expressed in terms of certain zeros $\al_1,\cdots,\al_n$ of
the function
\[\sin \pi s-\pi\sum c_{\om}\,(-\om)^{s-1}.\]
(The exponential is determined by taking $|\arg(-\om)|<\pi$.) The formulas are
\[a={1\ov n}\sum_{i=1}^n\al_i^2-{(n+1)\,(2n+1)\ov6},\]
\[ b={\prod_{|j|<n}G({j\over n}+1)^{n-|j|}\ov\prod_{i,j=1}^n
G({\al_i-\al_j\over n}+1)}.\]
Here $G$ is the Barnes $G$-function mentioned earlier.

To explain which zeros the $\al_i$ are, and the assumptions under which
this has been proved, we introduce a parameter $\la$ and the function
\bq\sin \pi s-\la\pi\sum c_{\om}\,(-\om)^{s-1}.\label{symb}\eq
Denote the zeros of this function by $\al_i(\la)\ (i\in\bZ)$, where $\al_i(0)=i$.
Then we take $\al_i=\al_i(1)$ for $i=1,\cdots,n$. Observe that the $\al_i(\la)$ are analytic 
functions of $\la$ away from those values for which the function has multiple zeros, and we
choose a path in $\bC$ from $\la=0$ to $\la=1$ which avoids all such values. Our
basic assumption is that there is a path in $\bC$ from $\la=0$ to $\la=1$ such that
everywhere on the path
\[\Re\,\al_i(\la)<\Re\,\al_{i+1}(\la),\ \ \ |\Re\,\al_i(\la)-i|<1.\]
(We believe that the second condition is redundant.) 

From these formulas we can compute the
asymptotics of our solutions (\ref{qk}) of (\ref{T}). They are given by
\[q_k(t)=A\log\Big({t\ov n}\Big)+\log B+o(1),\]
where for $k=1,\cdots, n$ the constants $A$ and $B$ are given by
\[A=2\,(\al_k-k),\ \ B=\prod_{1\leq j<k}{\Ga({\al_j-\al_k\over n}+1)
\ov\Ga({\al_k-\al_j\over n})}\prod_{k<j\leq n}{\Ga({\al_j-\al_k\over n})\ov
\Ga({\al_k-\al_j\over n}+1)},\]
and for other values of $k$ are given by periodicity.

In the special case $n=2$ this gives
the asymptotics for $\det(I\pm\la K)$, with $K$ given by (\ref{K}), and for the 
corresponding solutions of (\ref{sG}) for all $\la$ except real $\la$ satisfying 
$|\la|\geq1$. Consequently in this case at least the range of validity is the right one since the 
asymptotics are quite different for these $\la$ \cite{MTW}.\sp
\newpage
\noi{\bf 3. Outline of the derivation}\sp

If $R_{\la}$ denotes the resolvent kernel of $K$, the kernel of $K\,(I-\la K)\inv$, then
\[{d\ov d\la}\log\det(I-\la K)=-{\rm tr}\;R_{\la}.\]
Hence
\[\log\det(I-K)=-\int_0^1 {\rm tr}\;R_{\la}\,d\la=-\int_0^1d\la\int_0^{\iy}R_{\la}(x,x)\,dx,\]
where the path of integration can be taken to be that in our basic assumption. Now our
kernel is given by
\[K(x,y)=\sum c_{\om}\,{e^{-t\,[(1-\om)\,x+(1-\om\inv)\,x\inv]}\ov -\om x+y},\]
and we are looking for the asymptotics as $t\ra0$. Observe that as $t\ra0$,
\[e^{-t\,(1-\om\inv)\,x\inv}\ra1\ {\rm uniformly\ for}\ x\geq1,\ \ \ 
e^{-t\,(1-\om)\,x}\ra1\ {\rm uniformly\ for}\ x\leq1.\]
Therefore if we define kernels $K^{\pm}(x,y)$, still acting on $L^2(0,\iy)$, by
\[K^+(x,y)=\sum c_{\om}\,{e^{-t\,(1-\om)\,x}\ov -\om x+y},\ \ \ 
K^-(x,y)=\sum c_{\om}\,{e^{-t\,(1-\om\inv)\,x\inv}\ov -\om x+y},\]
with corresponding resolvent kernels $R^{\pm}_{\la}(x,y)$ then we might hope that
$R_{\la}(x,x)$ is close to $R^+_{\la}(x,x)$ when $x\geq1$ and close to $R^-_{\la}(x,x)$ when 
$x\leq1$. This is true, and using operator techniques it can be shown that
\[\int_0^{\iy}R_{\la}(x,x)\,dx=\int_1^{\iy}R^+_{\la}(x,x)\,dx+\int_0^1R^-_{\la}(x,x)\,dx
+o(1)\]
as $t\ra0$, uniformly in $\la$.
This is heart of the argument. Once we have this approximation then everything is plain 
sailing, since the operators $K^{\pm}$ can be transformed into Wiener-Hopf operators, and so 
there are exact integral representations for their resolvent kernels.

To describe this transformation on $K^+$, say, we observe that we may take $t=1$ since
the variable change $x\ra x/t$ replaces $K^+$ by the same kernel with $t=1$. Assuming this
to be the case we make the further assumption that each $\Re\,\om<0$. The reason we can
make this assumption
is that in the present discussion the $\om$ may be arbitrary complex numbers except
for nonnegative real numbers (when the kernels cease to represent trace class operators)
and that in the end everything will depend analytically on these $\om$. So that the formula
for the resolvent derived with the restriction holds generally.

Let us define
\[A(x,u)=\sum c_{\om}\,e^{-(1-\om)\,x}\,e^{\om xu},\ \ \
B(u,x)=e^{-xu}.\] These are both kernels of bounded operators from
$L^2(0,\iy)$ to $L^2(0,\iy)$ (here is where we use $\Re\,\om<0$) and
\[K^+(x,y)=\int_0^{\iy}A(x,u)\,B(u,y)\,du,\]
and so $K^+=AB$. Hence (our assumption guarantees that $I-\la K^+$ is invertible)
\[R^+_{\la}=AB(I-\la AB)\inv=A(I-\la BA)\inv B.\]
Therefore $R_{\la}$ is known once $(I-\la BA)\inv$ is. But the operator
$BA$ on $L^2(0,\iy)$ has kernel
\[BA(u,v)=\sum{c_{\om}\ov (u+1)-\om\,(v+1)},\]
and under the change of variable $u\ra e^{\xi}-1,\ v\ra e^{\eta}-1$ the
operator is transformed into the operator, also on $L^2(0,\iy)$, with
kernel
\[\sum {c_{\om}\ov e^{(\xi-\eta)/2}-\om\,e^{-(\xi-\eta)/2}}.\]
This is a Wiener-Hopf operator (its kernel is a function of the
difference of the  arguments) and so its resolvent is expressible in
terms of a factorization of its symbol, $1$ minus $\la$ times the Fourier
transform of the function
\[\sum {c_{\om}\ov e^{\xi/2}-\om\,e^{-\xi/2}}.\]
After a change of variable the Fourier transform becomes a Mellin
transform and the symbol becomes
\[1-{\la\pi\ov \sin \pi s}\sum c_{\om}\,(-\om)^{s-1}.\]
This is how the function (\ref{symb}) arises and why everything is 
expressed in terms of it.

The symbol is factored into products and quotients of $\Ga$-functions
whose arguments  involve the zeros $\al_i(\la)$. To see how the Barnes
function appears consider an integral of the form
\bq\int_{\hf-i\iy}^{\hf+i\iy}{\Ga'({s\ov n}+\be)\ov\Ga({s\ov n}+\be)}\,f(s)\,ds,\label
{int}\eq
where $f$ has period $n$. (Integrals like this which involve the $\al_i$
arise in the computation of the constant $b$.) To evaluate this we
integrate
\[{G'({s\ov n}+\be)\ov G({s\ov n}+\be)}\,f(s)\]
over the infinite rectangle with vertices $\hf\pm i\iy$ and $n+\hf\pm
i\iy$. This is equal to (\ref{int}) by the periodicity of $f$ and the
fact that
\[{G'({s+n\ov n}+\be)\ov G({s+n\ov n}+\be)}-{G'({s\ov n}+\be)\ov G({s\ov n}+\be)}=
{\Ga'({s\ov n}+\be)\ov\Ga({s\ov n}+\be)}.\]
On the other hand this is also equal to
\[2\pi i\sum_{\al}{G'({\al\ov n}+\be)\ov G({\al\ov n}+\be)}\;
{\rm res}\,(f,\al),\]
where $\al$ runs over the poles of $f$ inside the rectangle. In the
integrals that arise these poles are exactly the points
$\al_1,\cdots,\al_n$, and so (\ref{int}) is equal to
\[2\pi i\sum_{i=1}^n{G'({\al_i\ov n}+\be)\ov G({\al_i\ov n}+\be)}\;
{\rm res}\,(f,\al_i(\la)).\]\sp

\begin{center}{\bf Acknowledgements}\end{center}

The work of the first author was supported in part by the National
Science Foundation through Grant DMS--9303413, and the work of the second
author was supported in part by the National Science Foundation through
Grant DMS--9424292.


\begin{thebibliography}{4}

\bibitem{B} E.~W.~Barnes, {\it The theory of the $G$-function\/}, Quart.\
J.\  Pure and Appl.\ Math.\  {\bf 31} (1900) 264--314.

\bibitem{BT} E.~L.~Basor and  C.~A.~Tracy, {\it Asymptotics of a tau
function and Toeplitz determinants with singular generating
functions\/},  Int.\ J.\ Mod.\ Phys.\ {\bf A7} (1992) 83--107. 

\bibitem{BLC} D.~Bernard and A.~LeClair, {\it Differential equations for
sine-Gordon correlation functions at the free fermion point\/}. Nucl.\
Phys.\ {\bf B426} [FS] (1994) 534--558.

\bibitem{DIZ} P.~Deift, A.~Its and X.~Zhou, {\it A Riemann-Hilbert
approach to asymptotic problems arising in the theory of random matrix
models, and also in the theory of integrable statistical mechanics\/},
Ann.\ of Math.\ {\bf 146} (1997) 149--235.

\bibitem{D} F.~J.~Dyson, {\it Fredholm determinants and inverse
scattering problems\/}, Commun.\ Math.\ Phys.\ {\bf 47} (1976) 171--183.

\bibitem{G} M.~Gaudin, {\it Sur la loi limite de l'espacement des valeurs
propres d'une matrice al\'eatorie\/}, Nucl.\ Phys.\ {\bf 25} (1961)
447--458. [Reprinted in C.~E.~Porter, {\em Statistical Theory of Spectra:
Fluctuations\/}, New York, Academic Press, 1965.]

\bibitem{J} M.~Jimbo, {\it Monodromy problem and the boundary condition
for some Painlev\'e equations\/}, Publ.\ RIMS, Kyoto Univ.\ {\bf 18}
(1982) 1137--1161.

\bibitem{JMMS} M.~Jimbo,  T.~Miwa, Y.~M\^ori and M.~Sato, {\it Density
matrix of an impenetrable Bose gas and the fifth Painlev\'e
transcendent\/}, Physica {\bf 1D} (1980) 80--158.

\bibitem{K} A.~V.~Kitaev, {\it Method of isometric deformation for
``degenerate'' third Painlev\'e equation\/},  J.\ Soviet Math.\ {\bf 46}
(1989) 2077--2082. 

\bibitem{MTW} B.~M.~McCoy, C.~A.~Tracy and T.~T.~Wu, {\it Painlev\'e
functions  of the third kind\/}, J.\ Math. Phys.\  {\bf 18} (1977)
1058--1092. 

\bibitem{SMJ} M.~Sato, T.~Miwa and M.~Jimbo, {\it Holonomic quantum fields, III, IV\/},
Publ. RIMS, Kyoto Univ. {\bf 15} (1979) 577--629, 871--972.

\bibitem{TW1} C.~A.~Tracy and H.~Widom, {\it Level-spacing
distributions  and the Airy kernel\/}, Commun.\ Math.\ Phys.\ {\bf 159}
(1994) 151--174.

\bibitem{TW2} C.~A.~Tracy and H.~Widom, {\it Level-spacing distributions
and  the Bessel kernel\/}, Commun.\ Math.\ Phys.\ {\bf 161} (1994)
289--309.

\bibitem{TW3} C.~A.~Tracy and H.~Widom, {\it Fredholm determinants,
differential equations and  matrix models\/}, Commun.\ Math.\ Phys.\
{\bf 163} (1994) 38--72.

\bibitem{TW4} C.~A.~Tracy and H.~Widom, {\it Fredholm determinants and
the mKdV/sinh-Gordon  hierarchies\/}, Commun.\ Math.\ Phys.\ {\bf 179}
(1996) 1--10.

\bibitem{TW5} C.~A.~Tracy and H.~Widom, {\it Asymptotics of a class of
solutions to the  cylindrical Toda equations\/}, Commun.\ Math.\ Phys.\,
to appear.

\bibitem{W1} H.~Widom, {\it The strong Szeg\"o limit theorem for
circular arcs\/}, Ind.\ U.\  Math.\ J.\ {\bf 21} (1971) 277--283.

\bibitem{W2} H.~Widom, {\it The asymptotics of a continuous analogue
of   orthogonal polynomials\/}, J.\ Approx.\ Th.\ {\bf 76} (1994) 51--64.

\bibitem{W3} H.~Widom, {\it Asymptotics for the Fredholm determinant of
the  sine kernel on a  union of intervals\/},  Commun.\  Math.\ Phys.\
{\bf 171} (1995) 159--180.

\bibitem{W4} H.~Widom, {\it Some classes of solutions to the Toda
lattice hierarchy\/}, Commun.\  Math.\ Phys.\ {\bf 184} (1997) 653--667.


\end{thebibliography}
\end{document}